\documentclass[aps, prd, twocolumn, amsmath, floats,floatfix, superscriptaddress,
nofootinbib, showpacs,groupedaddress]{revtex4-1}

\usepackage{amssymb}
\usepackage{amsmath}
\usepackage{verbatim}
\usepackage{mathrsfs}
\usepackage{amsfonts}
\usepackage{latexsym}
\usepackage{epsfig}
\usepackage{color}
\usepackage{graphicx}
\usepackage{units}
\usepackage{slashbox}

\def\be{\begin{equation}}
\def\ee{\end{equation}}
\def\beq{\begin{eqnarray}}
\def\eeq{\end{eqnarray}}

\begin{document}

\renewcommand{\t}{\times}


\title{Rapid growth of superradiant instabilities for charged black holes in a cavity}  

 \author{Juan~Carlos~Degollado}\email{jcdaza@ua.pt}
 \author{Carlos~A.~R.~Herdeiro}\email{herdeiro@ua.pt}
 \author{Helgi Freyr R\'unarsson} \email{helgi.runarsson@ua.pt}
   \affiliation{
   Departamento de F\'\i sica da Universidade de Aveiro and I3N, 
   Campus de Santiago, 3810-183 Aveiro, Portugal.
 }


\date{May 2013}


\begin{abstract}  
Confined scalar fields, either by a mass term or by a mirror-like boundary condition, have unstable
modes in the background of a Kerr black hole. Assuming a time dependence as $e^{-i\omega t}$, the growth time-scale of these unstable modes is set
by the inverse of the (positive) imaginary part of the frequency, Im$(\omega)$, which reaches a maximum value
of the order of
Im$(\omega)M\sim 10^{-5}$, attained for a mirror-like boundary condition, where $M$ is the black hole mass.  In this paper we study the minimally
coupled
Klein-Gordon equation for a charged scalar field in the  background of a Reissner-Nordstr\"om black
hole and show that the unstable modes, due to a mirror-like boundary condition, can grow several
orders of magnitude faster than in the rotating case:  we have obtained modes with up to  
Im$(\omega)M\sim  0.07$. We provide an understanding, based on an analytic approximation, to why the
instability in the charged case has a smaller timescale than in the rotating case. This faster
growth, together with the spherical symmetry, makes the charged case a promising model for studies of the fully
non-linear development of  superradiant 
instabilities. 
\end{abstract}




\maketitle

\section{Introduction.}
\label{sec:introduction}

In classical relativistic gravity, black holes are observer independent space-time 
regions unable to communicate with their exterior \cite{Hawking:1973uf}. Thus, within this
description, information captured by black holes is trapped therein forever and cannot be
recovered by exterior observers. 
 
Given this picture it is intriguing, at first, to realise that there is a 
classical process through which energy can be extracted from a black hole:
\textit{superradiant scattering}.  In one form, this process amounts to the amplification
of waves impinging on a Kerr black hole, provided the frequency $\omega$ and azimuthal
quantum number $m$ of the wave modes obey the condition $\omega<m\Omega_+$, where
$\Omega_+$ is the angular velocity of the outer Kerr horizon
\cite{Bardeen:1972fi,Starobinsky:1973a,Press:1972zz}. The extraction of energy and
consequent decrease of the black hole mass $M$ is, however, necessarily accompanied by the
extraction of angular momentum and consequent decrease of the  black hole spin $J$. In
fact, it was shown by Christodoulou \cite{Christodoulou:1970wf} that the particle analogue
of this  process - the Penrose process \cite{Penrose:1969pc} - is irreversible,
subsequently realised to mean that the black hole area never decreases
\cite{Christodoulou:1972kt}. Finally, the identification between black hole area and
entropy \cite{Bekenstein:1973ur,Bardeen:1973gs} made clear that it is only rotational
energy that is being extracted from the black hole, not information.  
  
In another form, superradiant scattering amounts to the amplification of charged 
waves impinging on a Reissner-Nordstr\"om (RN) black hole, provided the frequency $\omega$
and the charge $q$ of the wave modes obey the condition $\omega<q\Phi_+$, where $\Phi_+$
is the electric potential of the outer Reissner-Nordstr\"om horizon
\cite{Bekenstein:1973mi}. The extraction of (Coulomb) energy and consequent decrease of
the black hole mass $M$ is, in this case, necessarily accompanied by the extraction of
charge and consequent decrease of the  black hole charge $Q$, such that, again, the
area/entropy of the RN black hole does not decrease.

The existence of superradiant modes can be converted into an instability of the 
background if a mechanism to trap these modes in a vicinity of the black hole is provided:
heuristically, these modes are then recurrently scattered off the black hole and
amplified, eventually producing a non-negligible back-reaction on the background. This
possibility, anticipated by Zel'dovich \cite{Zeldovich:1971}, was named \textit{black hole
bomb} by Press and Teukolsky \cite{Press:1972zz} and has been studied extensively in the
Kerr case within the linear analysis (see e.g.
\cite{Cardoso:2004nk,Berti:2009kk,Hod:2009cp,Rosa:2009ei,Pani:2012vp,Witek:2012tr,Dolan:2012yt}). One of the
outcomes of these studies is that the maximum growth rate for the instabilities is associated to
modes with an imaginary part of their frequency, Im$(\omega)$, of  Im$(\omega)M\sim  1.74\times
10^{-7}$ for massive fields and  Im$(\omega)M \sim 6\times 10^{-5}$ for mirror like boundary
conditions \cite{Cardoso:2004nk,Dolan:2012yt}. The growth time scale is set by the inverse of
Im$(\omega)$.

The unstable states found in the Kerr case are  localised in a potential well found
outside the potential barrier of the effective potential. The growth of such states can be
seen at linear level, but  a fully
non-linear study is required to address the end-point of this instability. It has been suggested
that such end-point is attained after an explosive event called \textit{bosenova} \cite{Yoshino:2012kn}.
Progress in understanding this endpoint has been making use, and will certainly require further use,
of fully non-linear numerical simulations \cite{Witek:2010qc,Cardoso:2012qm}.

Considerable less attention has been devoted to the charged case, perhaps due to the 
lack of astrophysical motivation. Moreover, the studies found in the literature
\cite{Furuhashi:2004jk,Hod:2012zz,Degollado:2013eqa} discard the possibility of an instability in
the asymptotically
flat case, since
an analysis of the effective potential shows no potential well  for
quasi-bound states compatible with the superradiance condition. This picture can be altered,
however, if the black hole is enclosed in a cavity. The purpose of this work is to show that
imposing a mirror-like boundary condition at some radial coordinate $r_m$, the superradiant
instability  occurs and can have a much shorter time scale than in the rotating case. 

Very slow instabilities prove challenging to follow numerically since the very small growth rate may
be masked by numerical errors or other physical effects; an example of the latter is discussed in
\cite{Witek:2012tr}. Thus, our result suggests that RN black holes inside a cavity provide a setup
that may facilitate the numerical study of the non-linear
development of superradiant instabilities, not only because of the shorter time scale
but also due to the spherical symmetry. Such non-linear study will certainly yield valuable lessons
for the more relevant, but harder, rotating case.

This paper is organised as follows. In Sec. \ref{sec:bound-states} we describe the basic setup for a
charged massive scalar field with a mirror-like boundary condition in the background of a RN black
hole. In Sec. \ref{results} we discuss the results for the imaginary part of the frequencies for
various values of the background and field parameters. In Sec. \ref{conclusions} we provide an
understanding to why the growth rate of instabilities in the charged case can become larger than in
the rotating case and discuss our results.

\section{Mirrored quasi-bound states.}
\label{sec:bound-states}

We shall consider a massive, charged scalar field, $\Psi$, with mass $\mu$ and charge $q$, 
propagating in the background of a Reissner-Nordstr\"om
black hole. As explained in the Introduction, in order to have superradiant scattering, the field
needs to be charged, thus making it natural to be massive as well, in view of the known fundamental particles. Written in Boyer-Lindquist
type
coordinates the line element of the background is  
\begin{equation}
ds^2=-f(r)dt^2+\frac{dr^2}{f(r)}+r^2(d\theta^2+\sin^2\theta d\phi^2) \ ,  
\end{equation}
where 
\be
f(r)=\frac{(r-r_+)(r-r_-)}{r^2} \ , \ \ r_\pm\equiv M\pm\sqrt{M^2-Q^2} \ .
\ee
In the linear regime, the dynamics of the scalar field is described by the wave equation
\begin{equation}
\left[(D^\nu-iqA^\nu)(D_\nu-iqA_\nu)-\mu^2\right]\Psi=0 \ ,
\label{eq:we}
\end{equation}
where the electric potential satisfies $A_\nu dx^\nu=-Q/r \, dt$. 

Setting $\Psi(t,r,\theta,\phi)= e^{-i \omega t}
\sum_{\ell,m}Y_\ell^m(\theta,\phi)R_{\ell}(r)/r$, with $Y_\ell^m(\theta,\phi)$ the spherical
harmonics, the radial equation for each mode can be written as:
\begin{eqnarray}
 \label{eq:radial}
f(r)\frac{d^2}{d r^2}R_{\ell}(r) + f'(r)\left(\frac{d}{dr}R_{\ell}(r) -
\frac{1}{r}R_{\ell}(r)\right) +\quad  \nonumber \\
\left(\frac{1}{f(r)}\left(\omega-\frac{qQ}{r}\right)^2
-\frac{\ell(\ell+1)}{r^2}-\mu^2\right)R_{\ell}(r) = 0 \ , \ \ \
\end{eqnarray}
where $f'(r)= {df(r)}/{dr}$ and $-\ell(\ell+1)$ is the eigenvalue of the angular
operator. This equation can be written in a Schr\"odinger-like
form by
applying a transformation of coordinates $r=r(r^*)$
\begin{equation}
\label{eq:Slike}
\left[-\frac{d^2}{dr^{*2}}+V(r)\right]R_{\ell}(r)=\omega^2R_{\ell}(r) \ , 
\end{equation}
where $r^*$ is the Regge-Wheeler tortoise coordinate defined by $dr^*=dr/f(r)$.
The effective potential is given by 
\begin{equation}
 \label{eq:Veff}
V(r)= \frac{2qQ\omega}{r}-\frac{q^2Q^2}{r^2}+f(r)\left(
\frac{l(l+1)}{r^2}+\mu^2+\frac{f'(r)}{r} \right) \ .
\end{equation}
The fact that the effective potential depends on the unknown $\omega$ makes it
unorthodox as compared to standard potentials in  Schr\"odinger-like problems. Some information,
nevertheless, can be obtained by studying this unorthodox potential. In particular,
it has been shown in \cite{Hod:2012zz} that quasi-bound states of a massive charged scalar fields in
an extreme RN geometry do not contain superradiant states. It has been proven that the conditions
for $(i)$
the effective potential to have a well and $(ii)$ the frequency to obey $\omega<\omega_{c}\equiv
q\Phi_+$, cannot
be satisfied simultaneously. For the non-extremal RN geometry, on the other hand, it was shown in 
\cite{Furuhashi:2004jk}, using a matching technique, that the condition 
$qQ<M\mu$ is necessary for the field to satisfy the regular boundary conditions at infinity. 
But, as pointed out in the same reference, this condition is not satisfied by superradiant states. The previous inequality is the Newton-Coulomb requirement 
for the gravitational force to exceed the
electrostatic force, therefore naturally associated to a condition for bound states.

The evidence presented in the previous paragraph points out that quasi-bound states for a massive
charged scalar field in a RN
background cannot be in the superradiant regime. There is, however, a way to obtain superradiant
quasi-bound states in this background. The point is that the frequencies of the quasi-bound
states are determined by both the parameters of the system and by the boundary conditions; thus changing the latter may allow bound states to obey the superradiant condition. 

The standard boundary behaviour for the quasi-bound states of a massive scalar field that can extend to asymptotic infinity is to decay exponentially; this follows from the fact that the effective potential tends asymptotically to the mass, generating a well.
If a mirror is placed at some radial
coordinate $r_m$ outside the black hole, on the other hand, the outer boundary condition is modified
such as the field vanishes exactly at $r_m$ and its
proper frequencies become determined by the position of the mirror \cite{Press:1972zz,
Cardoso:2004nk,
Dolan:2007mj}.
Since one can place the mirror at arbitrary positions, the scalar field might have frequencies that
are in the superradiant regime. As we show in the next section, this is indeed the case, and most
interestingly, the value of the time scale of the instability 
for the charged black hole can become considerably shorter than in the rotating counterpart of this
problem.

In order to compute the spectrum of bound states, we found it more convenient to numerically 
integrate the radial equation \eqref{eq:radial} imposing the appropriate boundary conditions.
In the vicinity of the horizon $r_{+}$, we impose an ingoing wave-like condition  \cite{Press:1972zz,
Cardoso:2004nk}
\begin{equation}
\label{eq:closeh}
R_{\ell}(r)\sim e^{-i (\omega-\omega_c)r^{*}}\ .
\end{equation}
The outer boundary condition is determined by the position of the mirror $r_{m}$. At this radius 
the mirrored states must vanish and hence $R_{\ell}(r_{m})=0$. 
The algorithm to find the frequencies is then the following: 
we start integrating the radial equation with the behaviour given
by \eqref{eq:closeh} outward from
$r=r_{+}(1+\varepsilon)$ - in the calculations presented in the following section we used
typically $\varepsilon\sim 10^{-8}$ - with an arbitrary value of $\omega$ and stop the integration at
the radius of the mirror. This procedure gives us a value
for the wave function at $r_m$, as a function
of the frequency. 
The integration is repeated varying the frequency until 
$R_{\ell}(r_m)=0$ is reached with the desired precision, thus obtaining the frequency of the
mirrored state. 
%

\section{Results}
\label{results}
We shall now exhibit the behaviour of the imaginary part of the frequency 
(also de real part in Fig. \ref{fig:freq-plots}) 
as a function of the mirror radius $r_m$, for various values of $q,\mu$ and $Q$. The black hole mass
is set to $M=1$, corresponding to a choice of scale. All the modes displayed
correspond to
$\ell=1$, since these are the modes for which the instability is expected to be stronger
\cite{Detweiler:1980uk,Furuhashi:2004jk}.

\begin{widetext}

\begin{figure}[h!]
\vspace{0.2cm}
\includegraphics[width=0.4\textwidth]{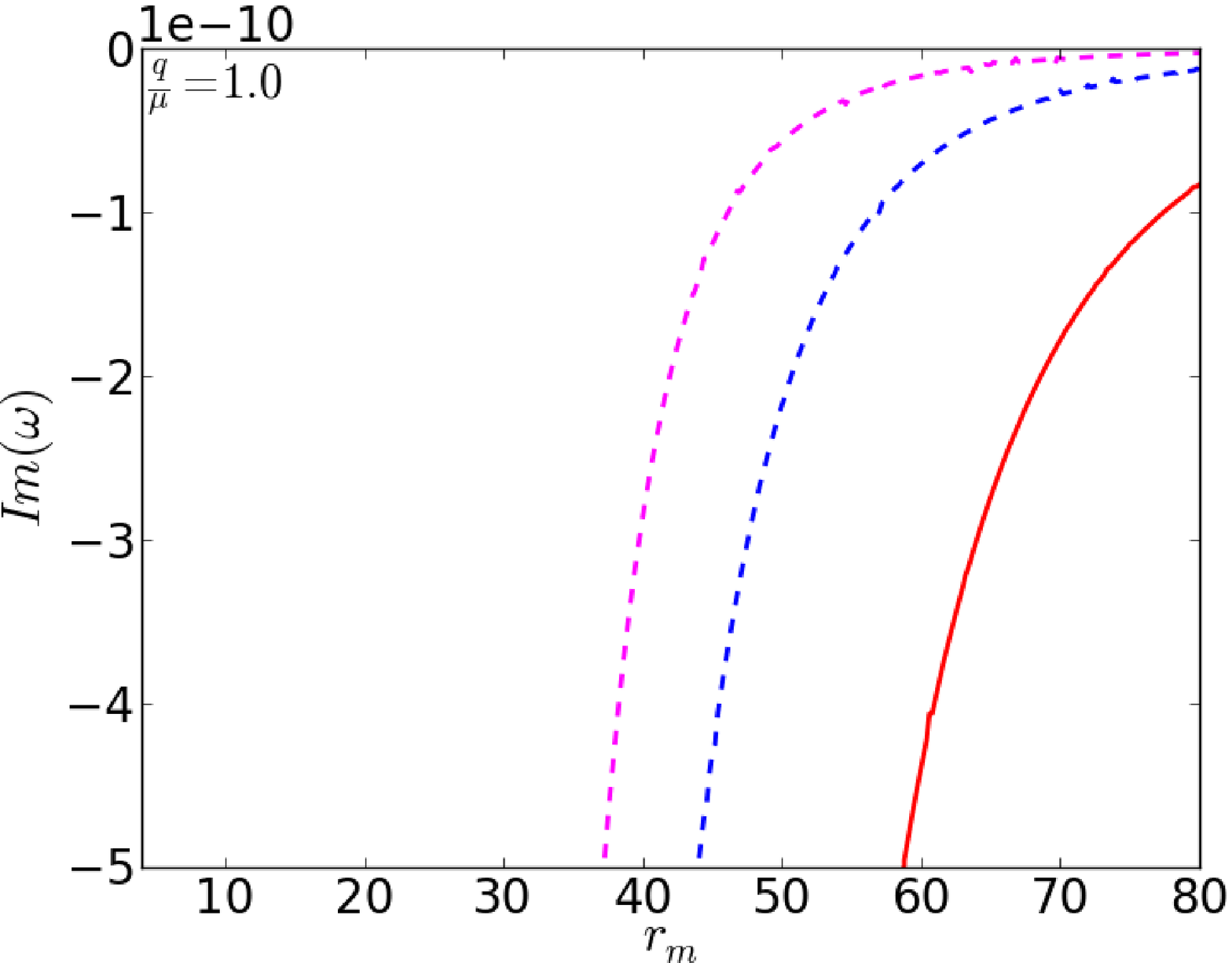} \,  
\includegraphics[width=0.4\textwidth]{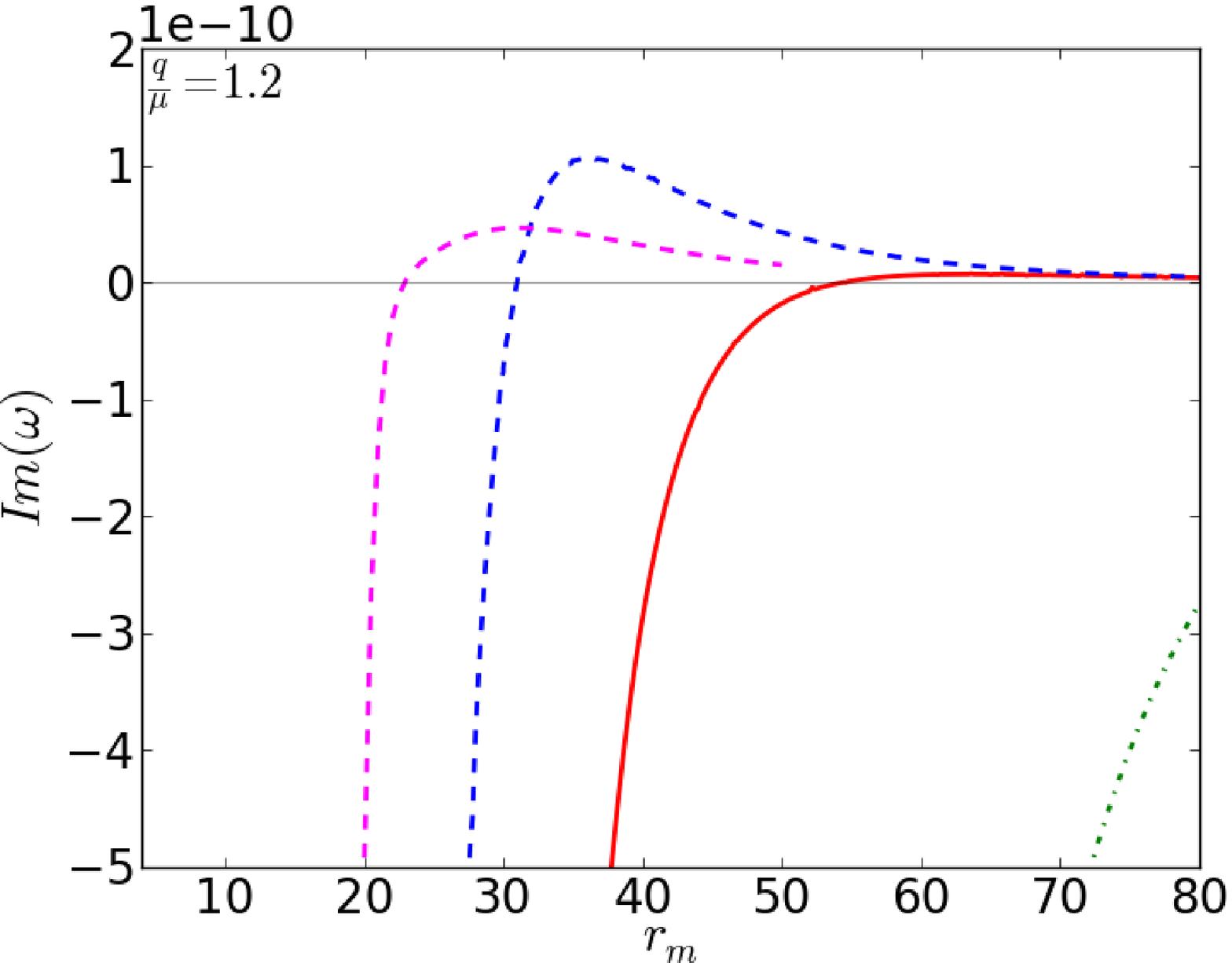}
\includegraphics[width=0.4\textwidth]{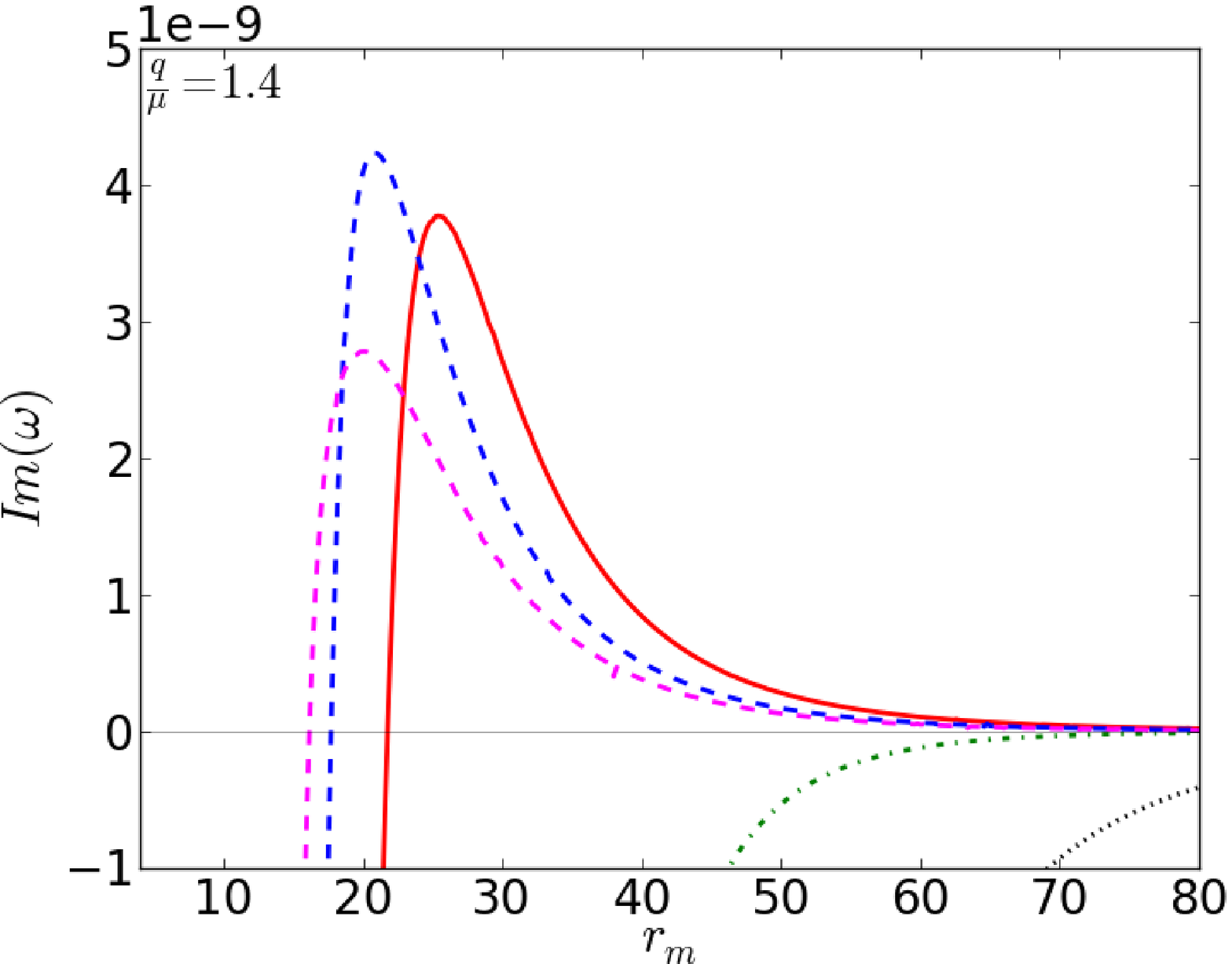} \,
\includegraphics[width=0.4\textwidth]{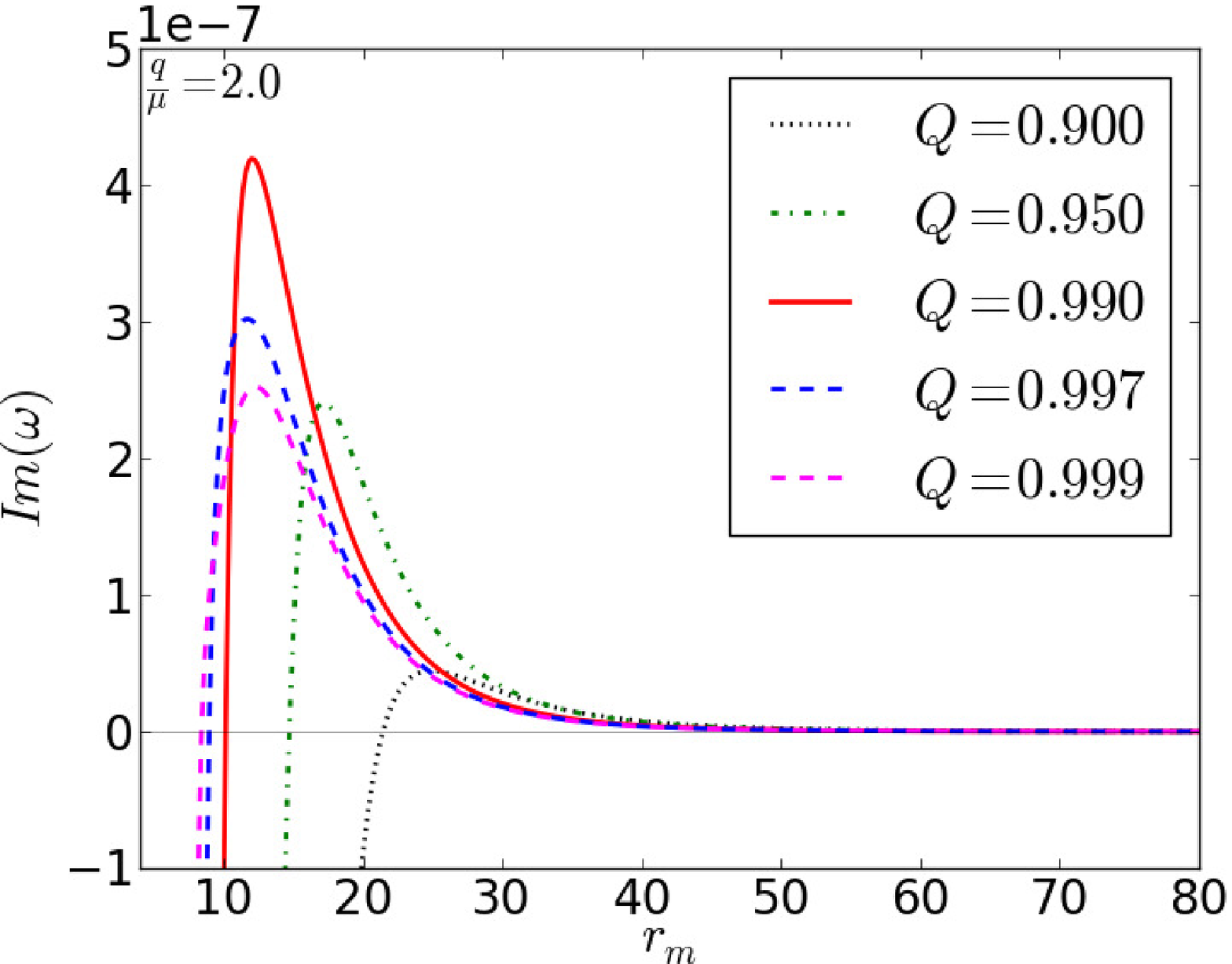}
 \caption{The imaginary part of the frequency plotted versus the radius of the mirror for various
ratios of the scalar charge, $q$ to scalar mass $\mu$. The scalar mass is $\mu=0.3$. We included the line Im$(\omega)=0$ to help visualizing where the curves have Im$(\omega)>0$.}
\label{fig:freq-plots-nolog}
\end{figure}

\end{widetext}

In Fig. \ref{fig:freq-plots-nolog} we show the imaginary part of the frequency as a function of the
mirror radius for different values of the ratio $q/\mu$ and $Q$.
We see that when the ratio is unity, no amount of black hole charge yields Im$(\omega)>0$; that is, there are no superradiant modes.
As the ratio increases, however, mirror radii greater than some minimum value will have positive
imaginary parts for a given $Q$. These facts are compatible with the following interpretations.
Firstly, superradiant modes have a
maximum frequency, thus they will have a minimum wavelength and hence a minimum radius for the
mirror is required for obtaining Im$(\omega)>0$.  Secondly, an analysis of the area
formula for RN
black holes reveals that an increase of area requires the small quantities of charge $dQ$ and mass
$dM$ extracted to be in a ratio greater than unity. Indeed, if the area $A=4\pi r_{+}^2$ increases by a small quantity $dA$, then requiring $dA>0$ yields
\begin{equation}
 dA=\frac{\partial A}{\partial M}dM+\frac{\partial A}{\partial Q}dQ>0 \quad \Rightarrow \quad
\frac{dQ}{dM}>1\ .
\end{equation}
Since our goal is to analyse how large the
instability may become we shall focus, in the following, on values for the scalar charge and mass
where their ratio is larger than unity.

In Fig.~\ref{fig:freq-plots} and  Fig.~\ref{fig:freq-plots-Q} we fix, respectively, the field charge
$q$ and the field mass $\mu$. From these figures we observe that as the scalar mass (scalar charge)
increases the magnitude
of the imaginary part of the frequency decreases (increases). A generic observation is that the real part of the frequencies
approaches the numerical value of the field's mass  and the imaginary part decreases monotonically
as $r_m$ increases.
Moreover, from
the latter figure  we see that if the scalar charge is increased, the black hole charge which gives
the maximum imaginary part of the frequency
increases and eventually becomes the extremal case.

\begin{widetext}

\begin{figure}[ht]
\vspace{0.2cm}
\includegraphics[width=0.33\textwidth]{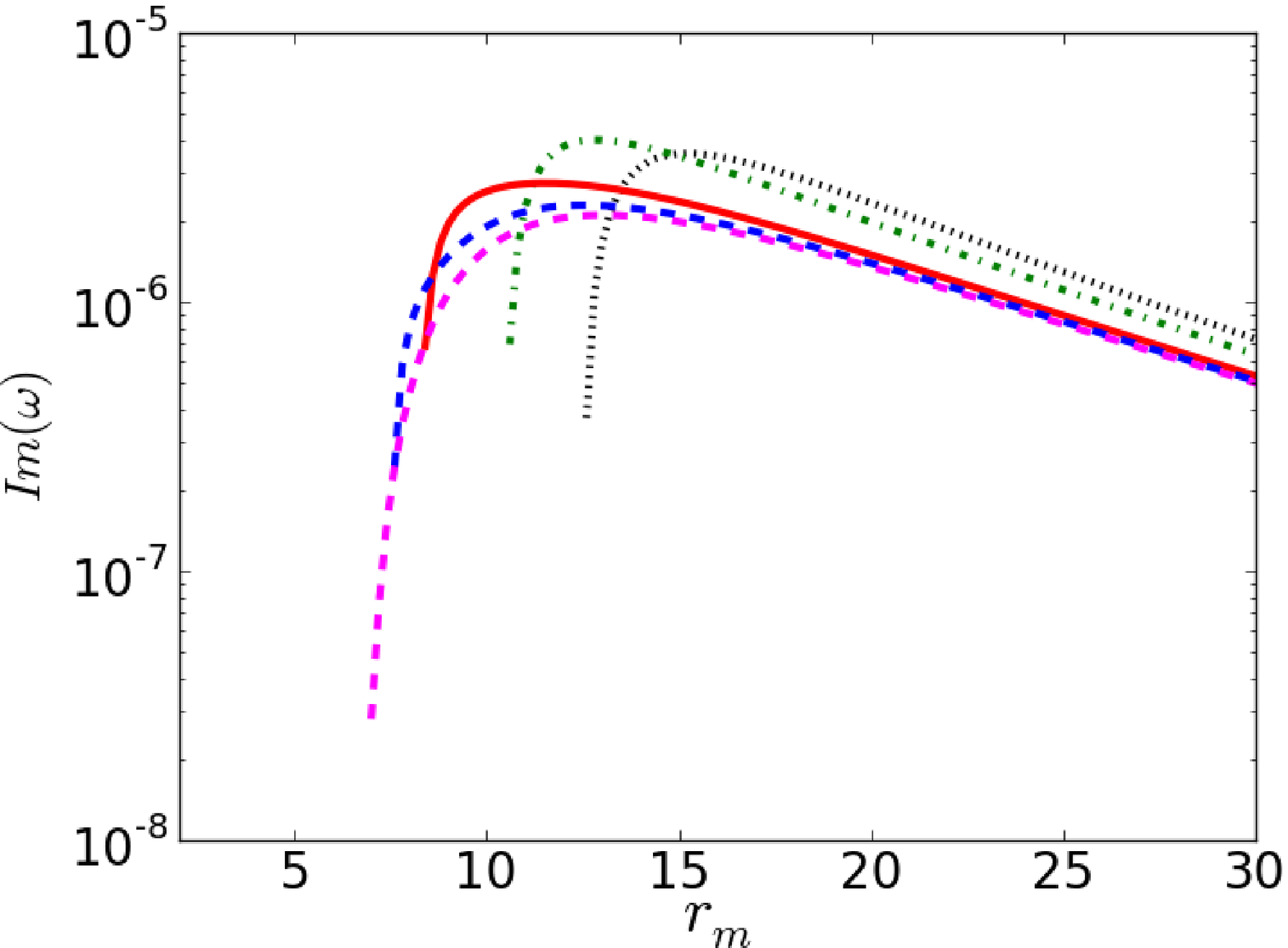} \,  
\includegraphics[width=0.31\textwidth]{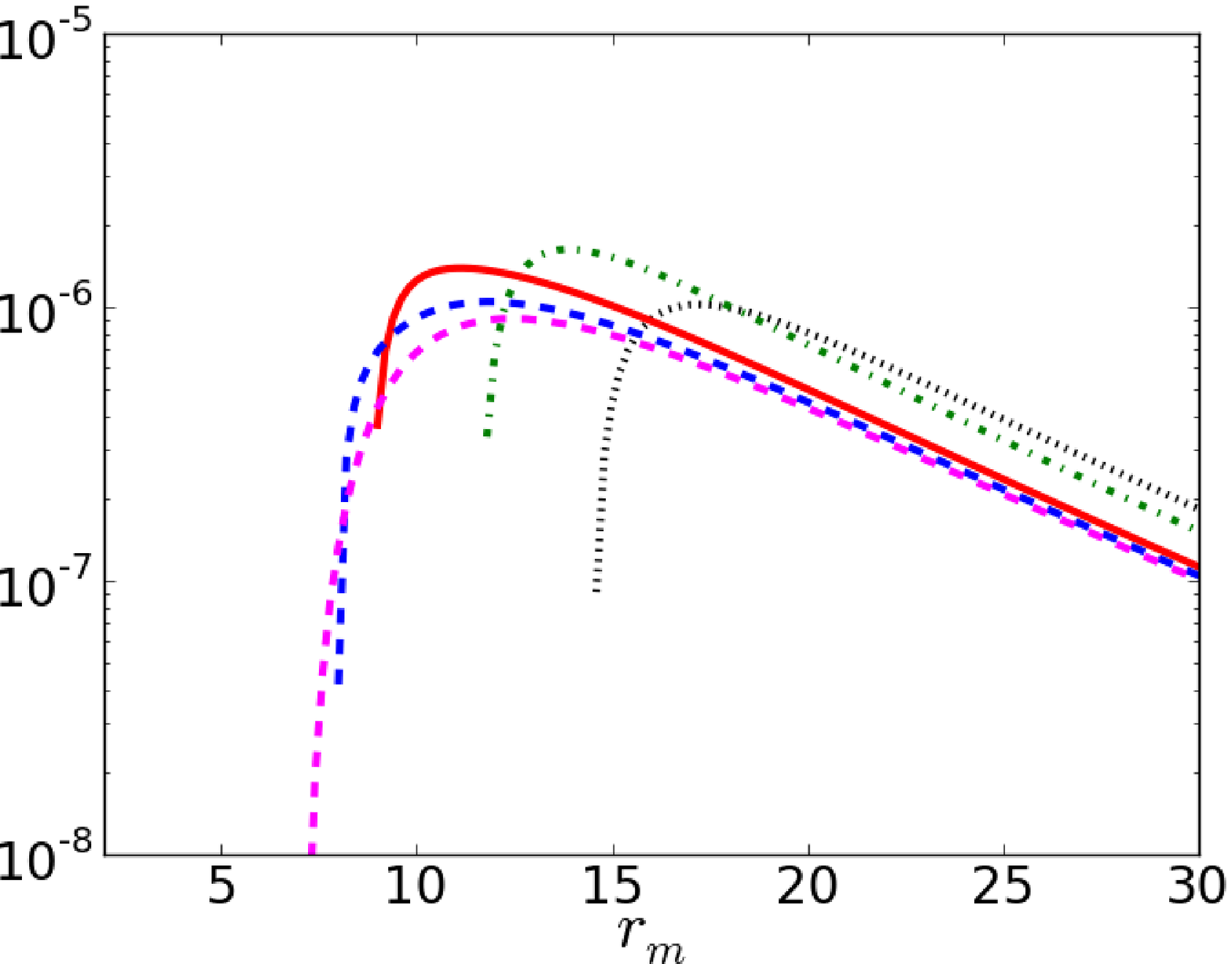}\,
\includegraphics[width=0.31\textwidth]{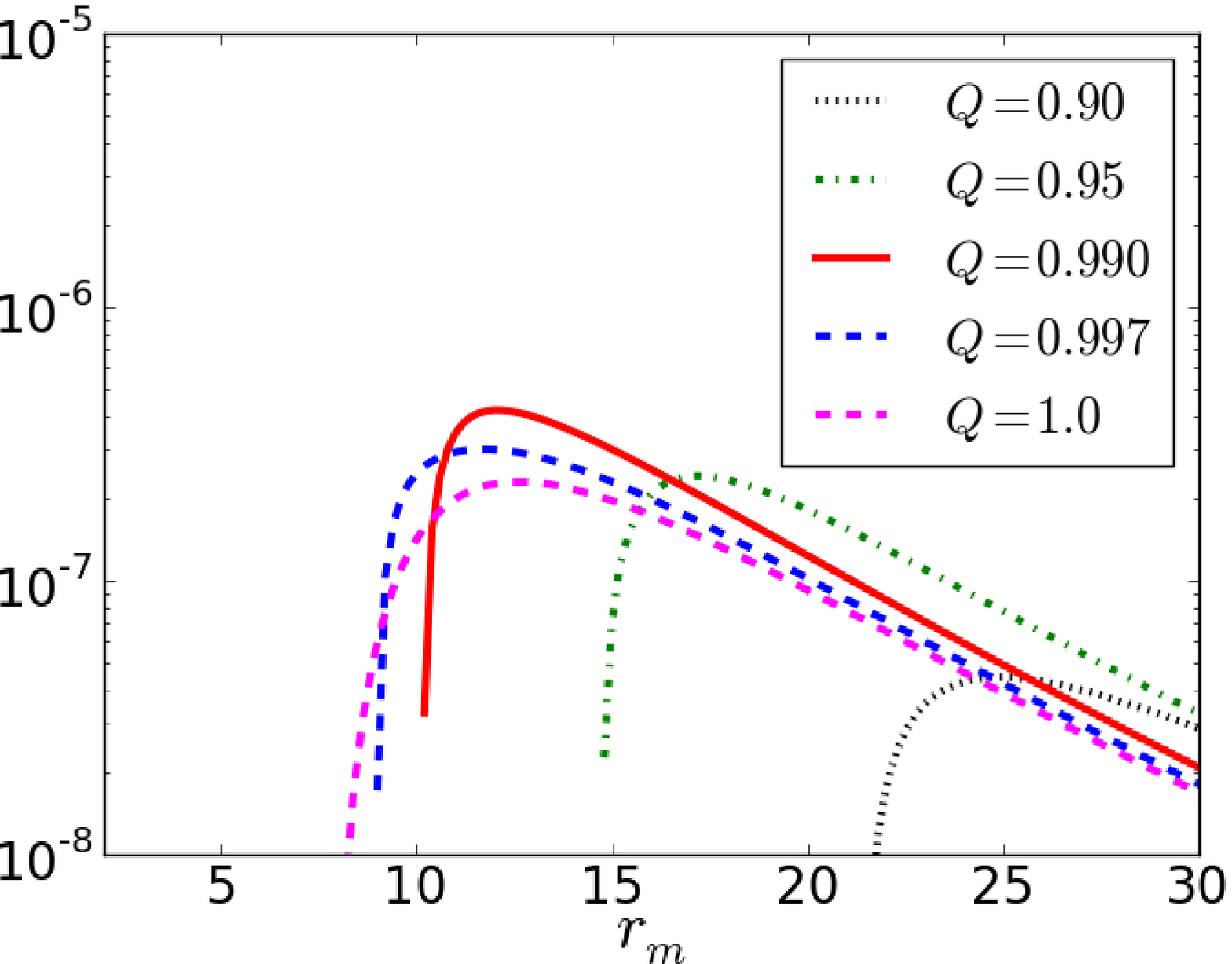} 
\includegraphics[width=0.33\textwidth]{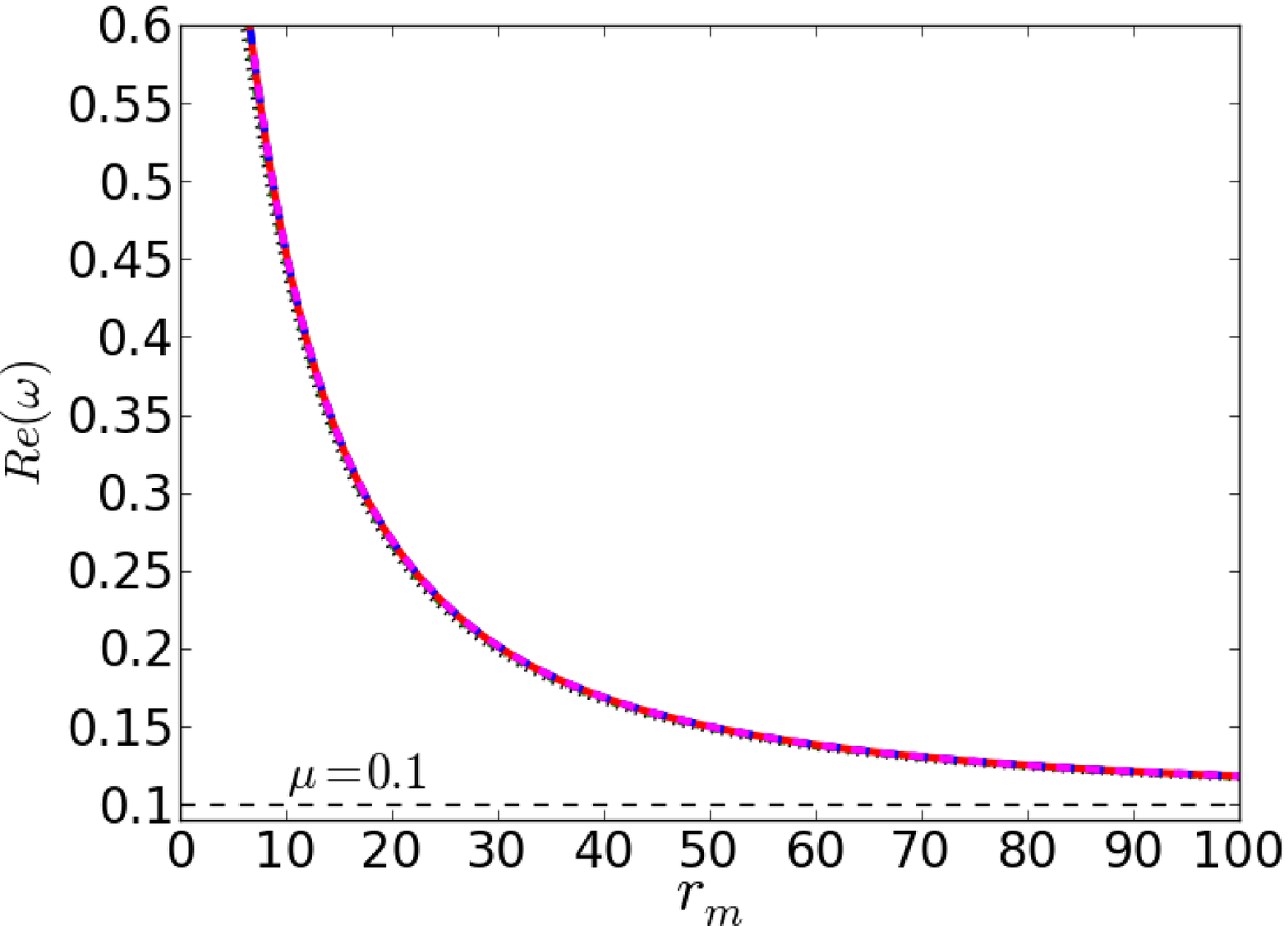}\,
\includegraphics[width=0.31\textwidth]{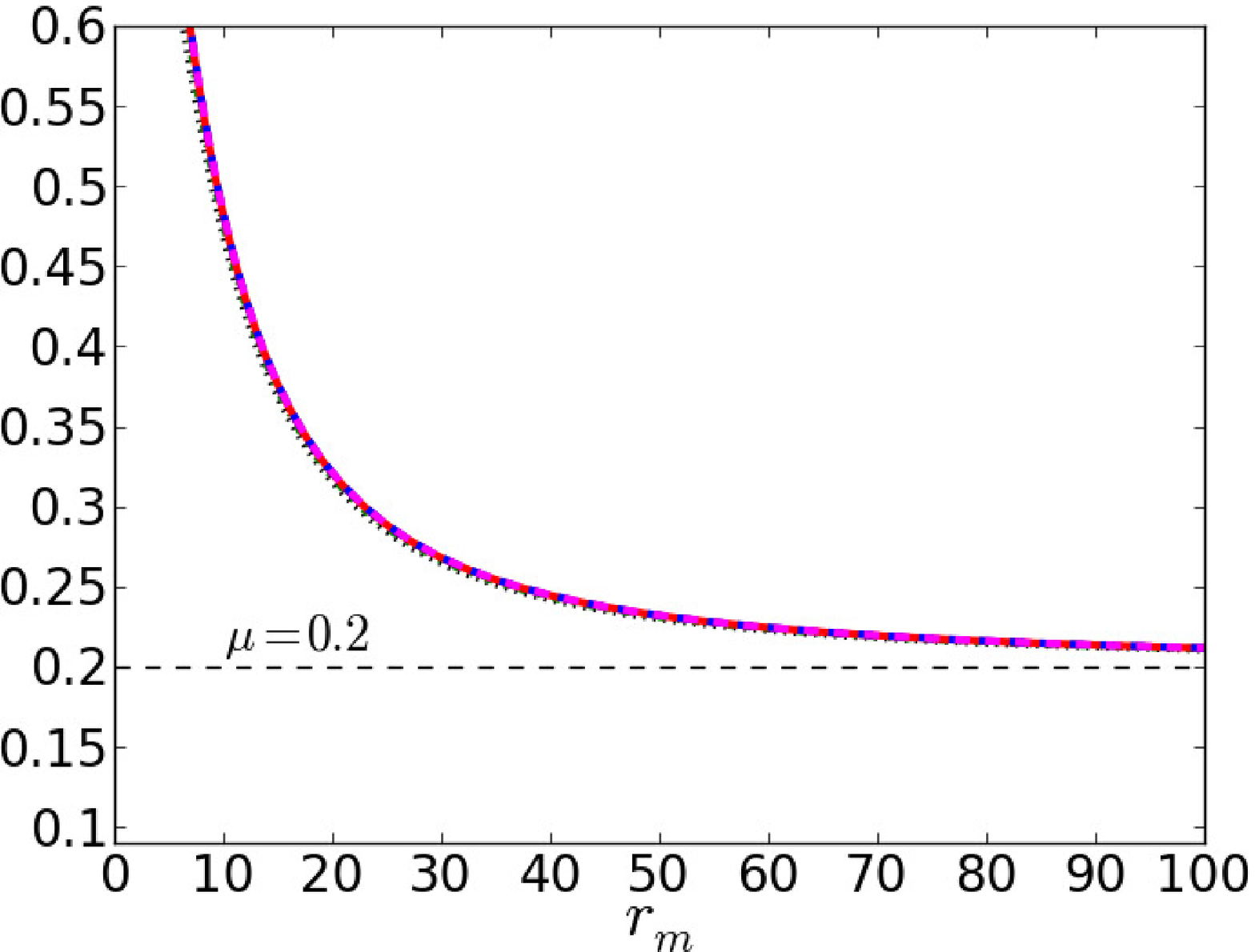}\,
\includegraphics[width=0.31\textwidth]{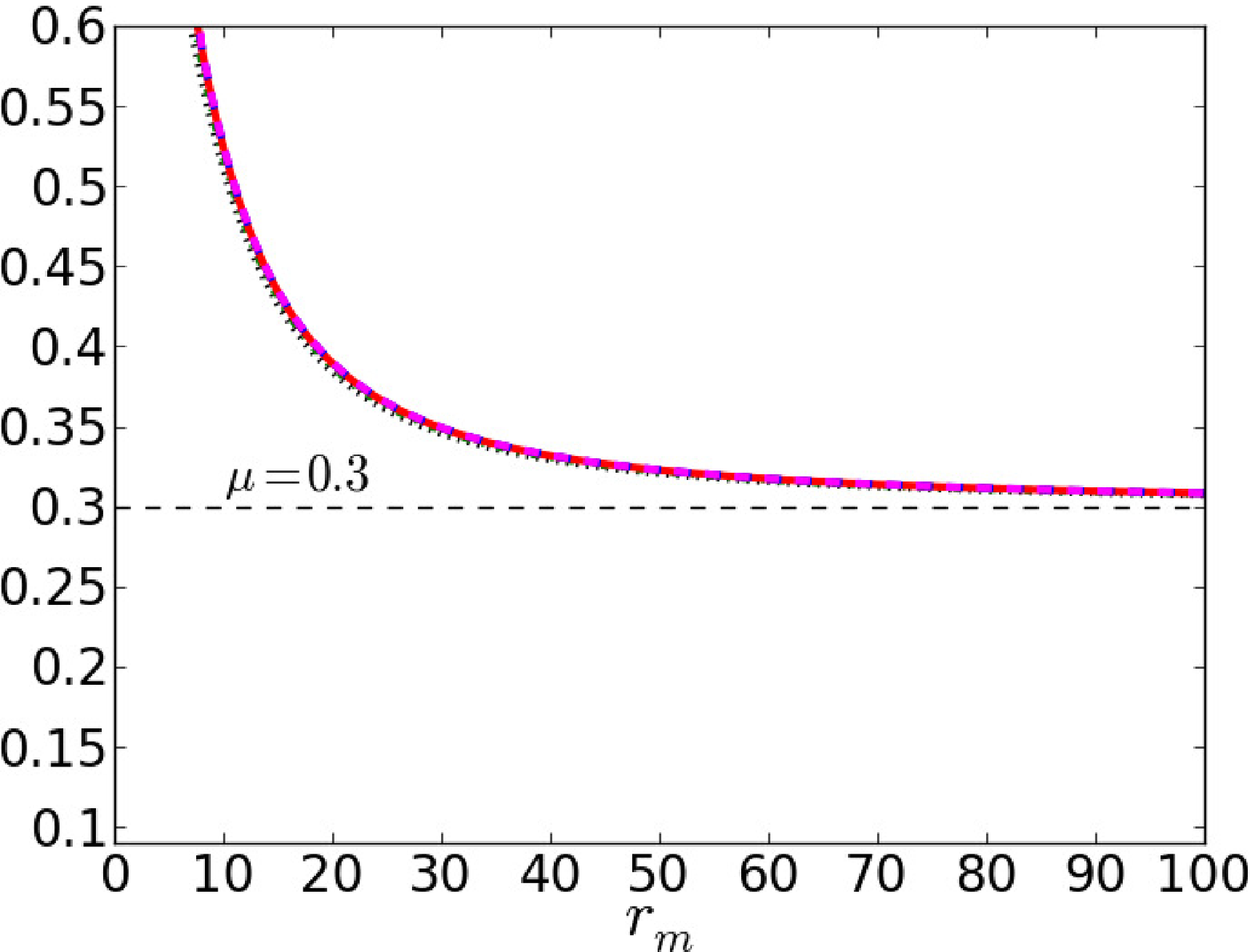}
 \caption{The imaginary and real part of $\omega$ drawn as a function of the mirror radius $r_m$ for
various values of the black hole charge, $Q$, and the scalar mass, $\mu$: $\mu=0.1,0.2,0.3$ for the
left, middle and right column respectively. We took $q=0.6$.}
\label{fig:freq-plots}
\end{figure}

\begin{figure}[ht]
\vspace{0.2cm}
\includegraphics[width=0.33\textwidth]{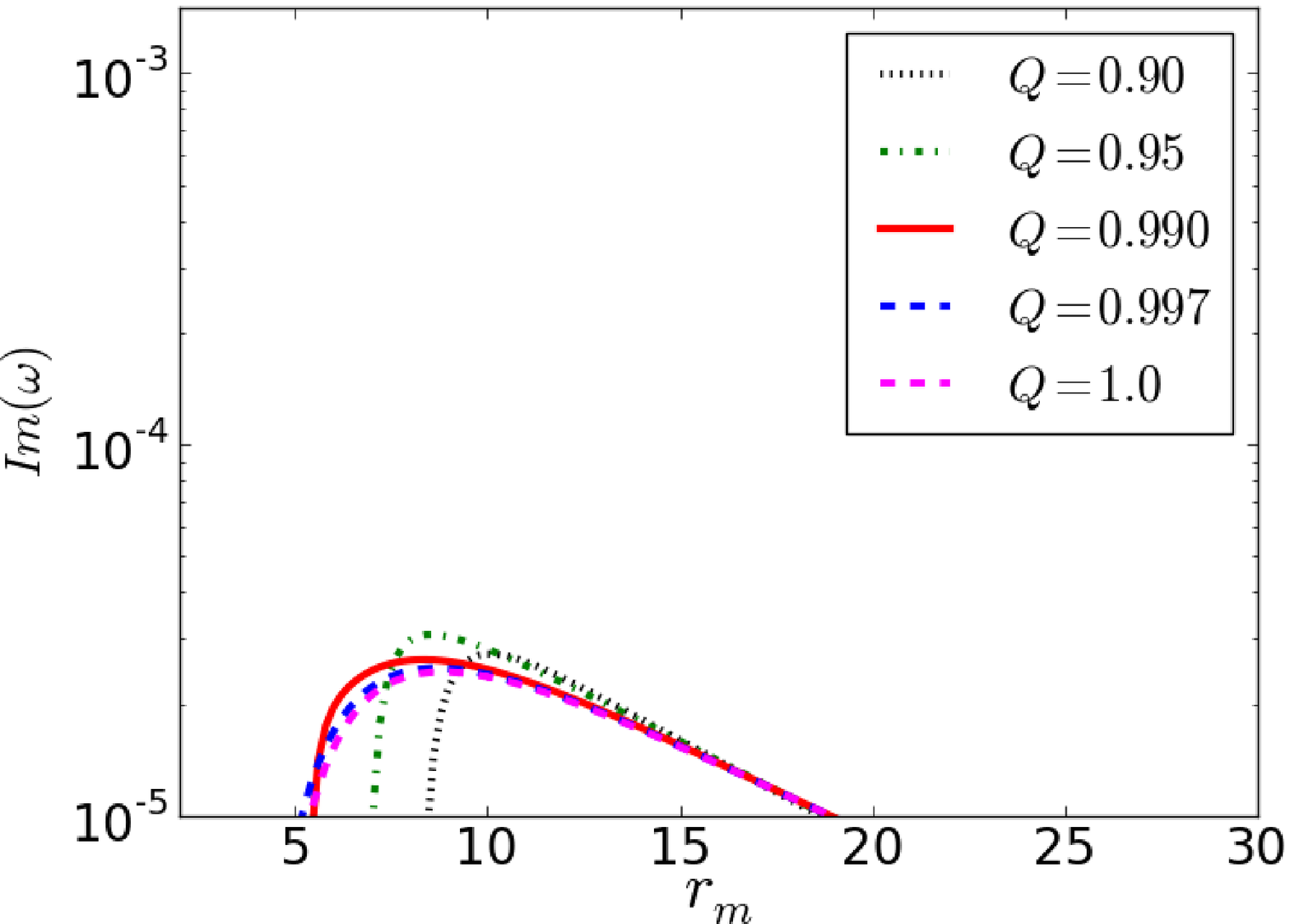} \,  
\includegraphics[width=0.31\textwidth]{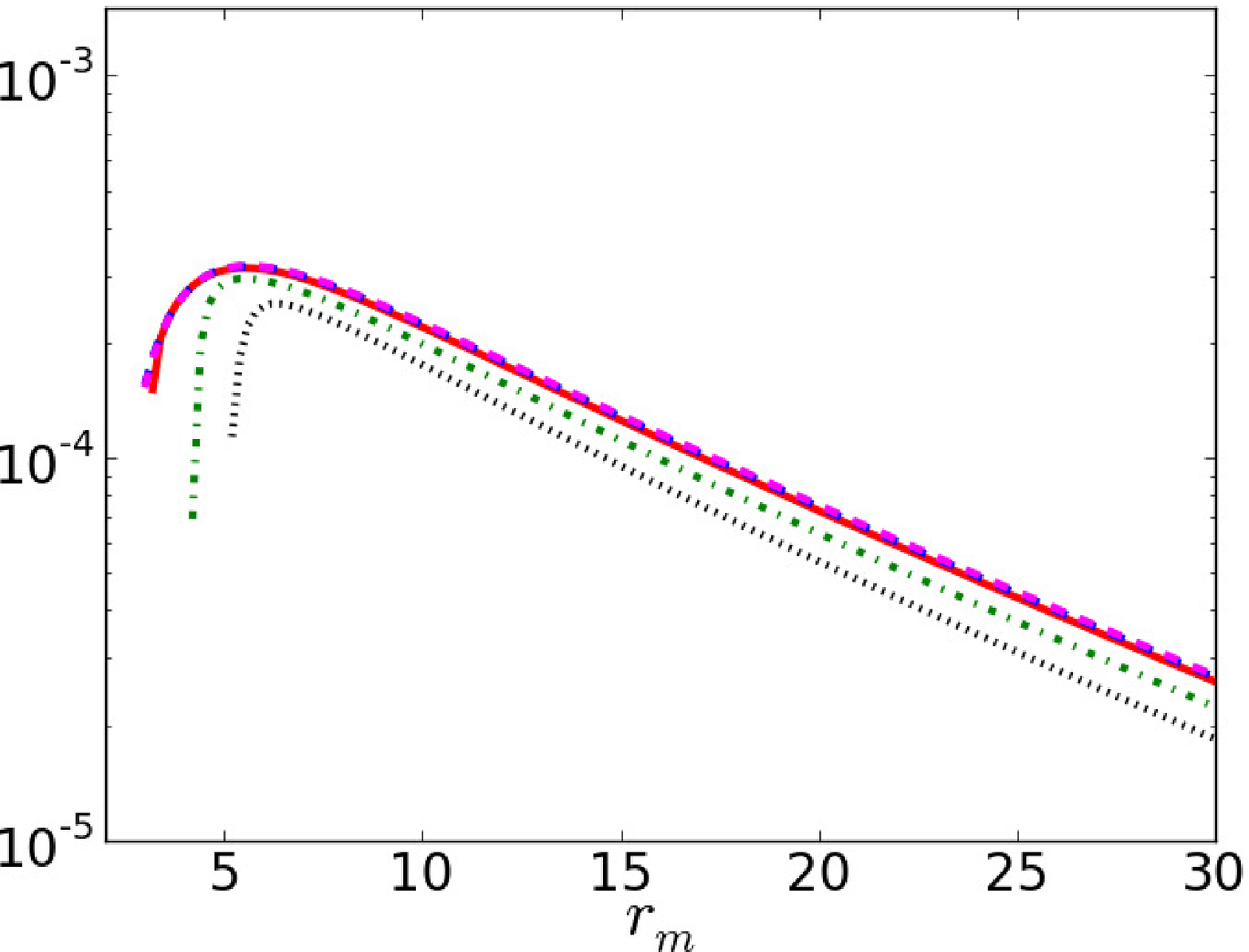}\,
\includegraphics[width=0.31\textwidth]{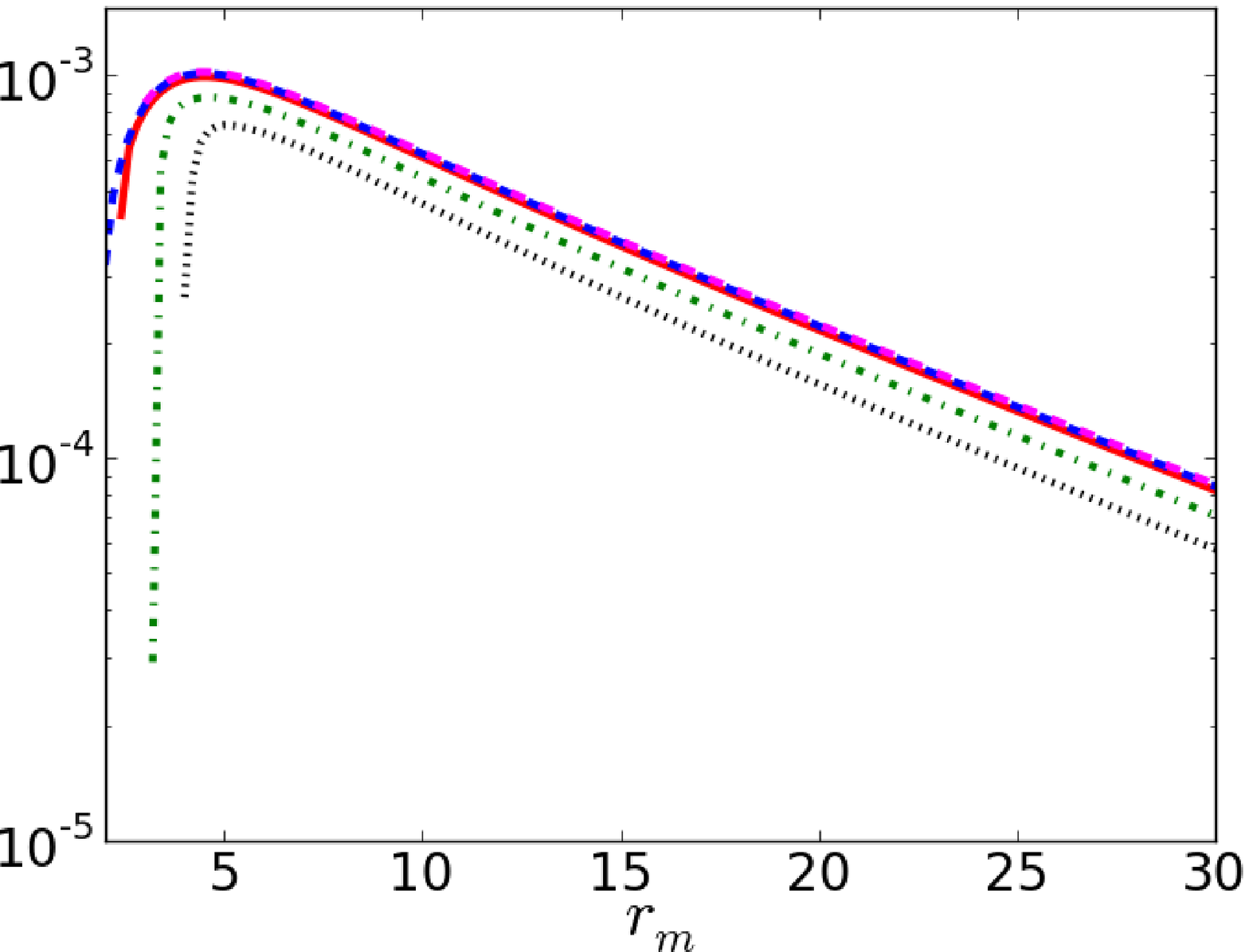} \
 \caption{The imaginary and real part of $\omega$ drawn as a function of the mirror radius $r_m$ for
various values of the black hole charge, $Q$, and the scalar charge, $q$: $q=0.9,1.5,2.0$ for the
left, middle and right column respectively. We took $\mu=0.1$.}
\label{fig:freq-plots-Q}
\end{figure}

\end{widetext}

\begin{figure}[h!]
\includegraphics[width=0.47\textwidth]{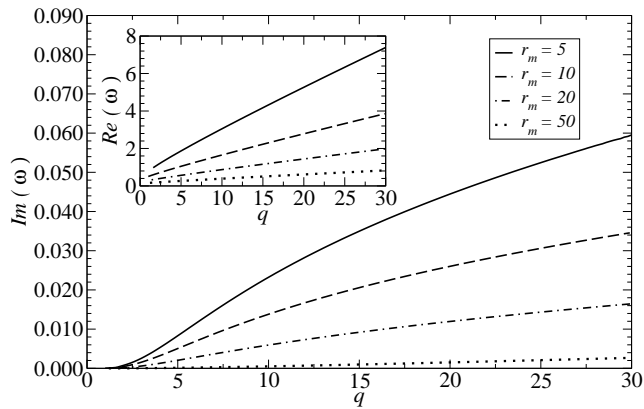}
\caption{The imaginary part (real part in the inset) of the frequency as a function of $q$ for
$Q=0.9$, $\mu=0.1$ and $r_m=5,10,20,50$.}
\label{fig:w_vs_q}
\end{figure}

These results indicate that in order to get the maximum
amplification of the scalar field, then $(i)$
the black hole should be extremal, or at least close to extremal, and $(ii)$ the scalar field should be
as
light as possible
but with the highest possible charge. This latter expectation is confirmed in Fig. \ref{fig:w_vs_q}, where
it is seen that fixing $Q$ and $r_m$ the imaginary part of the frequency grows monotonically with $q$ (as does
the real part).

These data suggests that the growth of Im$(\omega)$ with $q$ will continue. 
As $q$ increases, however, the integration of the radial equation becomes difficult. The first issue
arises because the coefficients of \eqref{eq:radial} might differ by several orders of magnitude. Close to the horizon, for instance, the last term in parenthesis of \eqref{eq:radial} can be
up to five orders of magnitude larger than the terms that multiply the derivatives. In this sense
the equation becomes \textit{stiff}. In order to ameliorate this difficulty we integrate the equation using
different methods and we report the values for the frequency at which both methods give the
same value\footnote{In most cases we used an Explicit Runge-Kutta integrator of 3th and 4th order;
however, for particular parameters we found necessary to use the Explicit Modified Midpoint method
provided by Mathematica \cite{mathematica}. }. 
Secondly, for large values of the frequency, the leading terms of $R_{\ell}(r)$ and
its derivative become very small close the horizon; with such small values the integrators find, very
frequently, the trivial solution $R_{\ell}(r)\equiv0$. For these reasons the largest imaginary
part we can quote is Im$(\omega)=0.07099\pm 0.0002$ for $q=40$, $Q=0.9$, $r_m=5.0$.



\section{Discussion and Conclusions}
\label{conclusions}
The main message in this paper is that for RN black holes enclosed in a cavity, superradiant
instabilities can be triggered by a charged scalar field; moreover, these instabilities can have a
considerably shorter timescale than the analogous problem in the Kerr background. 
A hint to why there is such a difference between rotating and charged black holes comes from
comparing the critical frequency for superradiance in both cases: $\omega_c=m\Omega_+$ for the
rotating and $\omega_c=q\Phi_+$ for the charged black holes. It follows that  in the charged case $q$
plays the same role that $m$ plays in the rotating case. But whereas the former is bounded by $\ell$, which
should be taken to be $\ell=1$ to maximise the instability 
\cite{Detweiler:1980uk,Furuhashi:2004jk}, there is no bound on $q$. Thus the critical frequency in a
fixed RN background can be made as large as one wishes by increasing $q$, thus rendering plausible
the existence of superradiant modes with very high frequency. 

The fact that $\omega_c$ grows with $q$ does not, however, lead to the conclusion that the
\textit{imaginary} part of the frequency should grow with $q$. But we can complement the above
argument with another one to make this point, as follows.

As we have shown with our numerical results, 
the smaller the value of the mass the greater the
value of the imaginary component of the frequency. In the limit of a zero mass field, an estimate
of the frequencies can be obtained analytically. The computation follows that in
\cite{Cardoso:2004nk}, where the Kerr black hole is considered and we provide only the main result. 

We shall now assume
that the Compton wavelength of the scalar particle is
larger than the typical size of the black hole, $1/\omega>>M$ (we have restored
the mass of the black hole for clarity).
Within this approximation it is possible to divide the space-time outside the horizon in two regions,
the near region, where $(r-r_{+})<<1/\omega$
and the far region, where  $(r-r_{+})>>M$. Then, one solves the wave equation in both regions
separately and where an
overlap occurs - in the region $M<<(r-r_{+})<<1/\omega$ - the solutions are matched.
Using this matching technique an approximation for the real part of frequency 
of the $n_{th}$ overtone $\omega_{n}$ in terms of the mirror radius can be obtained as   
${\rm Re}(\omega_{n}) = {j_{\ell+1/2,n}}/{r_{m}}$,
%
%
where $j_{\ell+1/2,n}$ is the $(n+1)^{th}$ root of the Bessel function of order $\ell$, $J_\ell$.
The imaginary part can be approximated as
\begin{equation}
{\rm Im}(\omega_{n})=-\gamma \frac{1}{r_{m}^{2(\ell+1)}}({\rm Re}(\omega_{n})-\omega_c) \ ,
\end{equation}
where
\begin{eqnarray}
\gamma=\frac{(-1)^{\ell}J_{-\ell-1/2}( j_{\ell+1/2,n} )}{J'_{-\ell-1/2}( j_{\ell+1/2,n} )}\left(
\frac{\ell!}{(2\ell-1)!!} \right)^2   \times
\\ \nonumber
\frac{r_{+}^2( M^2-Q^2 )^{\ell}}{ (2\ell)!(2\ell+1)!  }\frac{(2 j_{\ell+1/2,n}
)^{2\ell+1}}{2\ell+1} \prod_{k=1}^{l} (k^2+4\tilde\omega^2) \ ,
\end{eqnarray}
and $\tilde \omega = {r^{2}_{+}} \left(
{j_{\ell+1/2,n}}/{r_{m}}-\omega_{c}\right)/({r_{+}-r_{-}})$. The salient feature we wish to
emphasise is the dependence on $q$, which appears via the dependence on $\omega_c$. By inspection,
this suggests that Im$(\omega)$ grows with  $\omega_c$, and hence with $q$. This is indeed the
behaviour observed in the numerical results previously presented.

In table \ref{tab:num} we show some frequencies obtained by this analytic approximation and compare
them with the
values obtained with the numerical integration scheme previously presented. From the aforementioned
approximations one expects that lower frequencies yield a better analytical approximation. This is
indeed observed for the real part. An empirical observation is that the imaginary part is better
approximated by the analytical formula when the product $qQ$ is of the order of unity. This is also
seen for the frequencies displayed in the table. Thus, there are indeed regimes of
applicability for which the analytic approximation is legitimate.
\bigskip

\begin{center}
\begin{table}
  \begin{tabular}{ |c |c | c | c| c|}
    \hline
$q$ &$\omega_{Numerical}$ & $\omega_{Analytical}$ \\ 
\hline \hline
$0.1$ &$0.0452+4.7542\times10^{-10} i$ & $0.0449 +9.8835\times10^{-10} i$ \\
\hline
$1.2$ &$0.0605+7.1405\times10^{-7} i$ & $0.0449 +7.1534\times10^{-7} i$ \\
\hline
$1.6$ &$0.0657+3.8595\times10^{-6} i$ & $0.0449 +1.6754\times10^{-6} i$ \\
\hline
  \end{tabular}
\caption{Analytic and numerical frequencies for the mirrored states. The mirror is at $r_m=100$ and 
$Q=0.8$.}
\label{tab:num}
\end{table}
\end{center}
As already mentioned in the Introduction, the charged case we have studied herein does not seem to
have astrophysical relevance, mainly because if $Q/M\gtrsim10^{-13}(a/M)^{-1/2}(M/M_{\odot})^{1/2}$ the (Kerr-Newman) black hole is expected to discharge very quickly   \cite{Blandford:1977ds}. The interest of our study lies on
providing a setup wherein the non-linear
development of the superradiant instability might be more treatable. 

To have an idea of the orders of magnitude of the quantities used in our previous
computations we should convert them into physical units.
%
%
Taking the product $Mq\sim 30{\hbar c}/{G^{1/2}}$, that gives us the maximum value
for the imaginary part of the frequency obtained, 
for a black hole of one solar mass $M=M_{\odot}= 1.98\times 10^{33}g$, the charge of the scalar 
particle must be $q= 3.842\times 10^{-36}\, e$, and for a supermassive black hole of $M=
10^{8}M_{\odot}$ the particle must have a charge of $q=3.842\times 10^{-44}\, e$. Thus, realistic
particles will have values of $q$ much larger than those we have used, which, according to the trend
we have observed, should yield even lower timescales for the instability. 
Concerning the timescale of the instability,  the maximum value for the imaginary part of the
frequency is  Im$(\omega)M \sim
10^{-2}\frac{c^3}{G}$ which for 
a black hole of one solar mass  gives a time scale (e-folding time)
of 
$t=1/{\rm Im}(\omega)=4.924\times 10^{-4}\ s$, whereas for a supermassive black hole 
the e-folding time is  
$t=4.924\times 10^{4} s$. 

Finally, what will be the end state, in this setup, of the non-linear development of the
superradiant instability? Since the scalar field, after being amplified by the instability, cannot
leave the cavity, the end state is likely to be a scalar condensate around a charged black hole.
Observe that such scalar hair is not precluded by the usual theorems \cite{Mayo:1996mv}, since these
assume asymptotic flatness. This scenario parallels the fate of unstable charged black holes in
asymptotically Anti-de-Sitter space-times, against the condensation of a scalar field, which have
been of considerable interest over the last few years for studies of holographic superconductors
(see e.g. \cite{Hartnoll:2009sz}). Therefore, and in contrast to the asymptotically flat case,
Einstein-Maxwell-scalar field theory in a cavity, should have two families of spherically symmetric
solutions, at least for some range of the physical parameters. And the existence, in the linear
analysis, of a threshold mode with zero imaginary part, 
namely that with $\omega=\omega_c$ is, as in other well known cases such as the Gregory-Laflamme instability of cosmic strings \cite{Gregory:1993vy}, indicating such a branching in
the solutions of this theory.

\bigskip

\section*{Acknowledgements}
We would like to thank E. Radu for helpful discussions. JCD Acknowledges
CONACyT-M\'exico support. H.R was supported by an FCT grant from the
project PTDC/FIS/116625/2010.  This work was also supported by the {\it NRHEP--295189}
FP7-PEOPLE-2011-IRSES Grant.


\bibliographystyle{h-physrev4}
\bibliography{num-rel}

\end{document}